\documentclass[prl,preprint,%twocolumn,
superscriptaddress,showpacs,showkeys,amsmath,amssymb]{revtex4}

\usepackage{color,graphicx,overpic}
\usepackage{bm}
\usepackage{url}

\begin{document}

\title{Inherent Frequency and Spatial Decomposition of the Lorenz Chaotic Attractor}

\thanks{The first two authors, Gonzalo \'{A}lvarez and Shujun Li, are equivalent contributors of this letter.}
\author{Gonzalo \'{A}lvarez}
\email[E-mail: ]{gonzalo@iec.csic.es}%
\affiliation{Instituto de F\'{\i}sica Aplicada, Consejo Superior
de Investigaciones Cient\'{\i}ficas, Serrano 144---28006 Madrid,
Spain}

\author{Shujun Li}
\email[E-mail: ]{hooklee@mail.com}\homepage{www.hooklee.com}%
\affiliation{Department of Electronic Engineering, City University
of Hong Kong, Kowloon Toon, Hong Kong, China}

\author{Jinhu L\"{u}}
\affiliation{Institute of Systems Science, Academy of Mathematics
and Systems Science, Chinese Academy of Science, Beijing 100080,
China}

\author{Guanrong Chen}
\affiliation{Department of Electronic Engineering, City University
of Hong Kong, Kowloon Toon, Hong Kong, China}

\begin{abstract}
This letter suggests a new way to investigate 3-D chaos in spatial
and frequency domains simultaneously. After spatially decomposing
the Lorenz attractor into two separate scrolls with peaked spectra
and a 1-D discrete-time zero-crossing series with a wide-band
spectrum, it is found that the Lorenz chaotic attractor has an
inherent frequency uniquely determined by the three system
parameters. This result implies that chaos in the Lorenz attractor
is mainly exhibited when the trajectory crosses from one scroll to
another, not within the two scrolls. This is also true for some
other double-scroll Lorenz-like chaotic attractors, such as Chua's
attractor. Some possible applications of the inherent frequency
and the spatial decomposition are also discussed.
\end{abstract}



\maketitle

\newlength\figwidth
\setlength\figwidth{7.69cm} % this is the maximal size to make a 4-page format

Since the discovery of the first chaotic attractor by Lorenz
\cite{Lorenz63}, 3-D chaotic attractors have been extensively
studied to clarify how chaos occurs in the real world and to help
better understand the essence of chaos \cite{Sparrow:Lorenz1982}.
This letter suggests a novel way to investigate chaotic attractors
in frequency domain via a spatial decomposition method. It is
found that an inherent frequency, i.e., a prominent spectrum peak,
exists in all of the three variables of the Lorenz attractor (but
is particularly distinguished in the $z$-variable), and that the
Lorenz attractor can be spatially decomposed into two
single-scroll sub-attractors with peaked spectra at the inherent
frequency and a 1-D zero-crossing time series with wide-band
spectrum. This result strongly implies that, from the spectral
point of view, chaos in the Lorenz attractor is mainly exhibited
at the boundary of the two scrolls, not within them. Further
numerical experiments show that the inherent frequency is uniquely
determined by the three system parameters. The inherent frequency
and the spatial decomposition also exist for some other
Lorenz-like chaotic attractors with double scrolls, such as Chua's
attractor \cite{ChuaAttractor:JCSC1994}. Compared with previous
work on spectral analysis of ``phase incoherent chaos"
\cite{Farmer:PhaseCoherence:ANYAoS80}, this letter reveals much
subtler features inherently-existing in Lorenz-like chaotic
attractors.

\begin{figure}[!htbp]
\centering
\includegraphics[width=\figwidth]{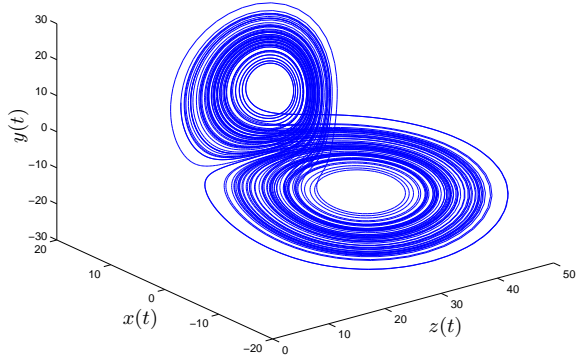}
\caption{The Lorenz attractor}\label{figure:Lorenz3D}
\end{figure}

Consider the following Lorenz system \cite{Lorenz63,
Sparrow:Lorenz1982}:
\begin{eqnarray}
\dot{x} & = & \sigma(y-x),\\
\dot{y} & = & rx-y-xz,\\
\dot{z} & = & xy-bz,
\end{eqnarray}
where $\sigma,r,b$ are three positive parameters. When
$\sigma=10,r=28,b=8/3$, the Lorenz system has a double-scroll
chaotic attractor shown in Fig. \ref{figure:Lorenz3D}. Because the
Lorenz system is invariant under the coordinate transformation
$(x,y,z)\to(-x,-y,z)$, the chaotic attractor is symmetric with
respect to the two planes $x=0$ and $y=0$. When $r>1$, the centers
of the two scrolls, which are two equilibrium points of the Lorenz
system, can be calculated as
$\left(\pm\sqrt{b(r-1)},\pm\sqrt{b(r-1)},r-1\right)$. Roughly
speaking, the double-scroll attractor is developed by the chaotic
trajectory spirally rotating around the two points in a chaotic
fashion.

Although the Lorenz chaotic attractor has been extensively studied
in the past three decades, the relationship between the chaotic
behaviors and the spectra of $x(t),y(t),z(t)$ has not yet been
clarified \cite{Farmer:PhaseCoherence:ANYAoS80}. This letter tries
to give a clearer description of this problem via a novel spatial
decomposition method. In the following, without loss of
generality, the system parameters are always set as the typical
values $\sigma=10,r=28,b=8/3$.

\begin{figure}[!htbp]
\centering
\begin{overpic}[width=\figwidth]{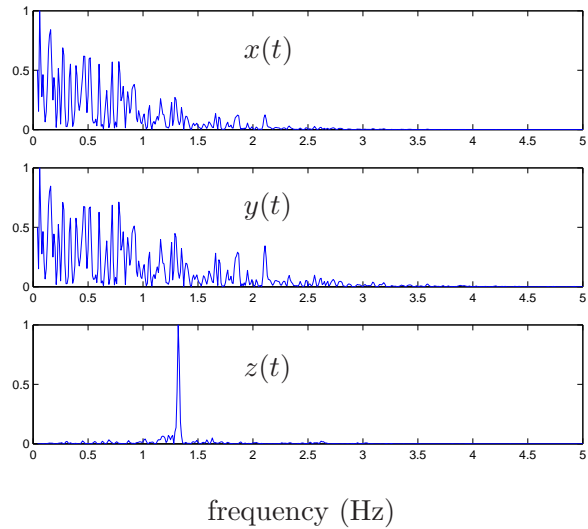}
    \put(40,70){$x(t)$}
    \put(40,43){$y(t)$}
    \put(40,15){$z(t)$}
\end{overpic}\\
frequency (Hz) \caption{The relative power spectra of $x(t)$,
$y(t)$ and $z(t)$}\label{figure:LorenzPSD}
\end{figure}

\begin{figure}[!htbp]
\centering
\includegraphics[width=\figwidth]{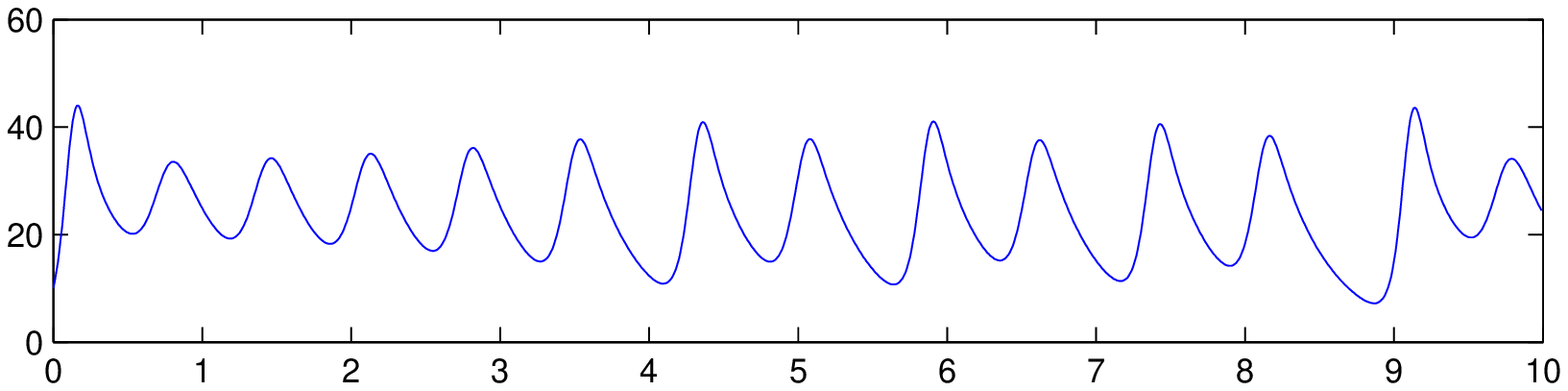}\\
time (sec) \caption{The nearly-periodic evolution of
$z(t)$}\label{figure:Lorenz_z}
\end{figure}

In Fig. \ref{figure:LorenzPSD}, the power spectra of $x(t)$,
$y(t)$ and $z(t)$ are shown (calculated by DFT with 4-term
Blackman-Harris window, the same hereinafter). It can be seen that
the spectra of $x(t)$ and $y(t)$ are locally wide-band in the low
frequency range and gradually approach to zero as the frequency
increases, which shows the existence of chaos in $x(t)$ and
$y(t)$. In comparison, the power spectrum of $z(t)$ is much
simpler: it has a significant spectral peak at the frequency
$f_z\approx 1.3$ Hz, which implies that $z(t)$ is nearly periodic,
i.e., weakly chaotic. Observing the time evolution of $z(t)$ shown
in Fig. \ref{figure:Lorenz_z}, one can see that the spectrum peak
of $z(t)$ is a natural reflection of the nearly-invariant
short-time period of $z(t)$.

Considering the spatial symmetry of the Lorenz attractor with
respect to the two planes $x=0$ and $y=0$, the spectrum peak of
$z(t)$ implies that the rotation frequency of the chaotic
trajectory around each scroll is nearly invariant and the mean
frequency is $f_z$. As a natural result, when the chaotic
trajectory rotates within each scroll, the local (i.e.,
short-time) spectra of $x(t)$ and $y(t)$ should also have a peak
at the frequency $f_z$. Is it really true? If so, then it can be
expected that the chaoticity of the Lorenz attractor is mainly
exhibited when the trajectory crosses the boundary between the two
scrolls.

\begin{figure}[!htbp]
\centering
\includegraphics[width=\figwidth]{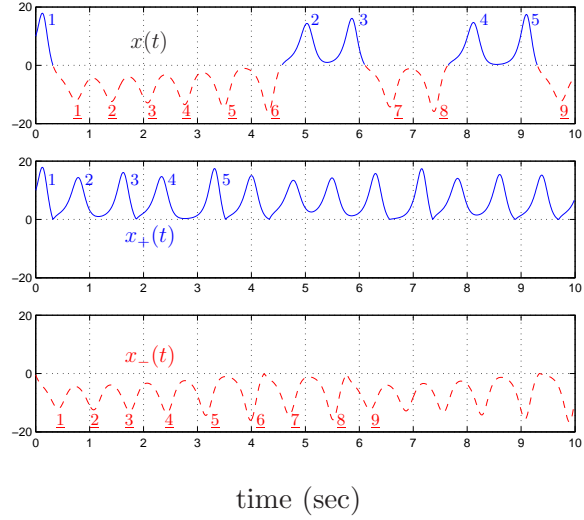}\\
time (sec) \caption{The extraction of $x_+(t)$ and $x_-(t)$ from
$x(t)$}\label{figure:Lorenz_x12}
\end{figure}

\begin{figure}[!htbp]
\centering
\includegraphics[width=\figwidth]{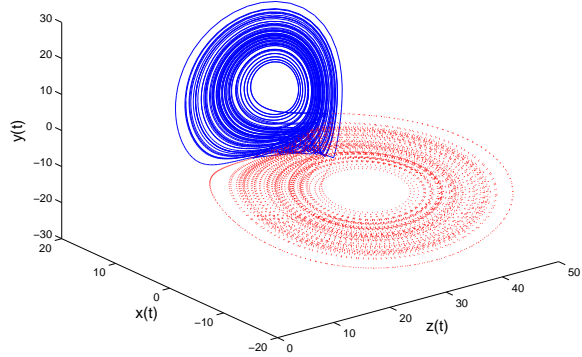}
\caption{The positive (solid) and the negative (dashed)
sub-attractors of the Lorenz attractor}\label{figure:Lorenz12}
\end{figure}

In the following, a simple spatial decomposition method is
introduced as a tool to give an answer. The basic idea of the
spatial decomposition method is to divide the chaotic trajectory
into three parts: two separate single-scroll sub-attractors and a
1-D zero-crossing time series. The method is described as follows:

\begin{itemize}
\item \textit{Two separate single-scroll sub-attractors}:

1) connect all disjoint positive parts of $x(t)$ together to make
an embedded signal $x_+(t)>0$, and connect all its disjoint
negative parts to make an embedded signal $x_-(t)<0$ (see Fig.
\ref{figure:Lorenz_x12} for a schematic illustration on how the
two embedded signals are extracted from the original signal
$x(t)$);

2) determine the sub-signals of $y(t),z(t)$ corresponding to
$x_+(t)$ and $x_-(t)$ in the time axis: $y_+(t),y_-(t)$ and
$z_+(t),z_-(t)$.

As a result, $(x_+(t),y_+(t),z_+(t))$ constitutes a positive
sub-attractor, and $(x_-(t),y_-(t),z_-(t))$ constitutes a negative
sub-attractor (see Fig. \ref{figure:Lorenz12}). Both
sub-attractors have a single scroll.

\item \textit{A 1-D zero-crossing time series
$\{ZC(i)\}_{i=1}^\infty$}:

1) use the zero-crossing times, i.e., the times when the chaotic
trajectory crosses the boundary $x=0$ from one scroll to another,
to create a discrete-time series $\{t_{zc}(i)\}_{i=1}^\infty$;

2) $ZC(i)$ is defined as the elapsed time between $t_{zc}(i+1)$
and $t_{zc}(i)$: $ZC(i)=t_{zc}(i+1)-t_{zc}(i)$.
\end{itemize}

\begin{figure}[!htbp]
\centering
\begin{overpic}[width=\figwidth]{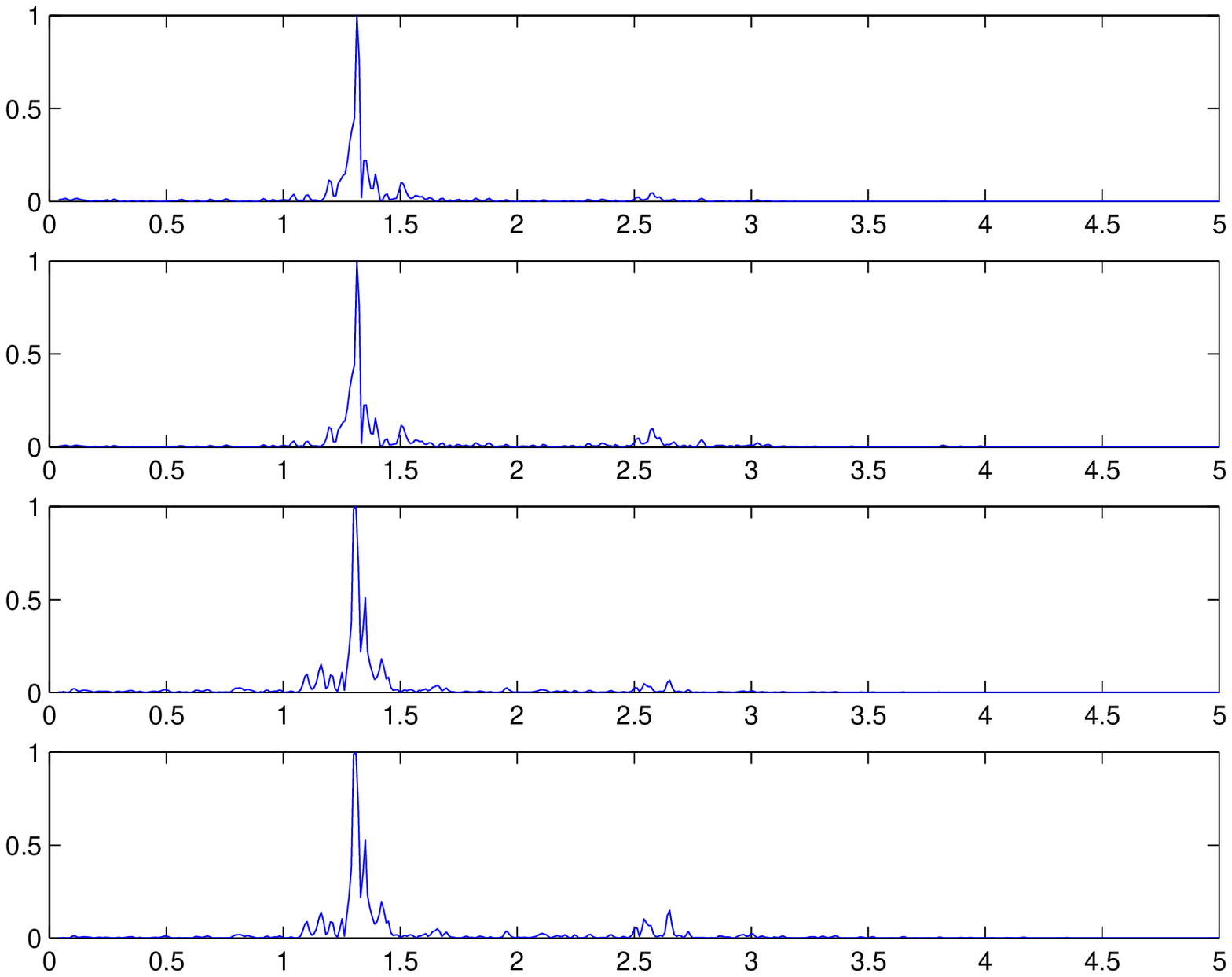}
    \put(45,12){$y_-(t)$}
    \put(45,30){$x_-(t)$}
    \put(45,50){$y_+(t)$}
    \put(45,70){$x_+(t)$}
\end{overpic}\\
frequency (Hz) \caption{The relative power spectra of $x_+(t)$,
$y_+(t)$, $x_-(t)$ and $y_-(t)$}\label{figure:Lorenz_psd_x1y1}
\end{figure}

Note that the above spatial decomposition can also be made with
respect to the plane $y=0$ to get two sub-attractors in the same
way. After the above spatial decomposition of the Lorenz
attractor, the three time series are analyzed via the spectral
analysis method. It can be expected that the spectra of the two
sub-attractors have a peak at the frequency of $f_z$ and the
spectrum of $ZC(i)$ is wide-band. The experimental results well
support this prediction. Observing the power spectra of $x_+(t)$,
$y_+(t)$, $x_-(t)$ and $y_-(t)$ shown in Fig.
\ref{figure:Lorenz_psd_x1y1}, it can be clearly seen that a
spectrum peak occurs at the same frequency $f_z\approx 1.3$Hz in
each spectrum. Note that the frequency $f_z$ of $z(t)$ is actually
a reflection of the nearly-periodic trajectory within the two
sub-attractors, since $z(t)$ is independent of the chaotic
zero-crossing behaviors.

\begin{figure}[!htbp]
\centering \begin{overpic}[width=\figwidth]{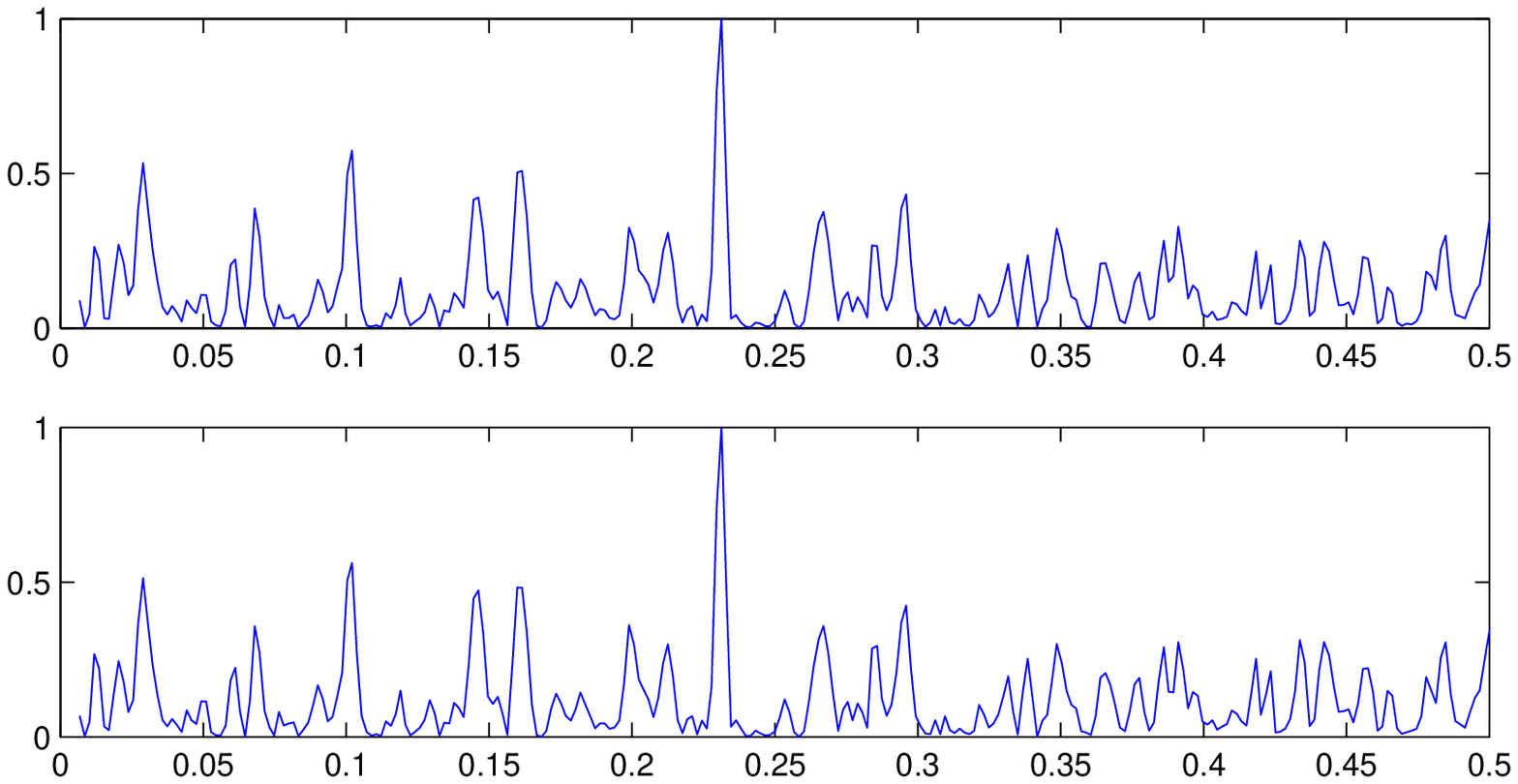}
    \put(60,45){a) $ZC(i)$}
    \put(60,18){b) $ZC_n(i)$}
\end{overpic}\\
frequency (the maximal frequency is 1Hz) \caption{The relative
power spectra of $ZC(i)$ and
$ZC_n(i)$}\label{figure:Lorenz_psd_ZC}
\end{figure}

Now, let us consider the power spectrum of $ZC(i)$ shown in Fig.
\ref{figure:Lorenz_psd_ZC}a, which keeps wide-band in the whole
frequency range and has many spectral peaks without decaying to
zero. Comparing all the five spectra, it is obvious that the local
wide-band spectra of $x(t)$ and $y(t)$ stems from the wide-band
spectrum of $ZC(i)$. Because the elapsed time between two
consecutive zero-crossing events is at least equal to the time of
one cycle of $x(t)$, i.e., the mean period of the Lorenz attractor
$T_z=1/f_z$, the occurrence frequency of the zero-crossing event
will not be greater than $f_z$. It is the reason why the original
spectra of $x(t)$ and $y(t)$ gradually decay to zero as the
frequency increases beyond $f_z$.

\begin{figure}[!htbp]
\centering \begin{overpic}[width=\figwidth]{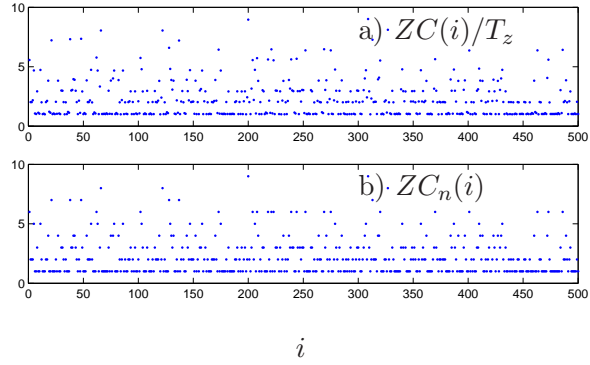}
    \put(60,45){a) $ZC(i)/T_z$}
    \put(60,18){b) $ZC_n(i)$}
\end{overpic}\\
$i$ \caption{$ZC(i)/T_z$ and $ZC_n(i)$}\label{figure:Lorenz_ZC}
\end{figure}

Here, note the following fact: since the two sub-attractors are
both approximately periodic, $ZC(i)$ will always be about $n$
times of $T_z$, where $n$ is an integer. In other words,
$ZC(i)/T_z$ approximates the number of rotational cycles of the
chaotic trajectory within the current sub-attractor. This can be
clearly seen from Fig. \ref{figure:Lorenz_ZC}a. Then, one can
define a new time series
$ZC_n(i)=\mbox{round}\left(ZC(i)/T_z\right)$, where
$\mbox{round}(\cdot)$ denotes the function rounding a real number
to a nearest integer. In Fig. \ref{figure:Lorenz_ZC}b, $ZC_n(i)$
is displayed to show its approximation to $ZC(i)/T_z$. Apparently,
$ZC_n(i)$ has a clearer physical meaning than $ZC(i)$, and has a
similar spectrum to $ZC(i)$. Figure \ref{figure:Lorenz_psd_ZC}
gives a comparison of the power spectra of $ZC(i)$ and $ZC_n(i)$,
which are almost identical.

The above spatial decomposition of the Lorenz attractor shows that
the chaotic trajectory's \textbf{short-time} behavior within each
scroll is approximately periodic (i.e., weakly chaotic) and that
the strongly chaotic behaviors mainly occur instantaneously when
the trajectory crosses the boundary of the two scrolls with a mean
invariant frequency not greater than $f_z$. Another obvious
physical meaning of the spatial decomposition is the collapse of
3-D chaos into 1-D space. Considering that any 3-D chaotic system
has only one positive Lyapunov exponent, such a collapse seems
apprehensible.

\begin{figure}[!htbp]
\centering
\includegraphics[width=\figwidth]{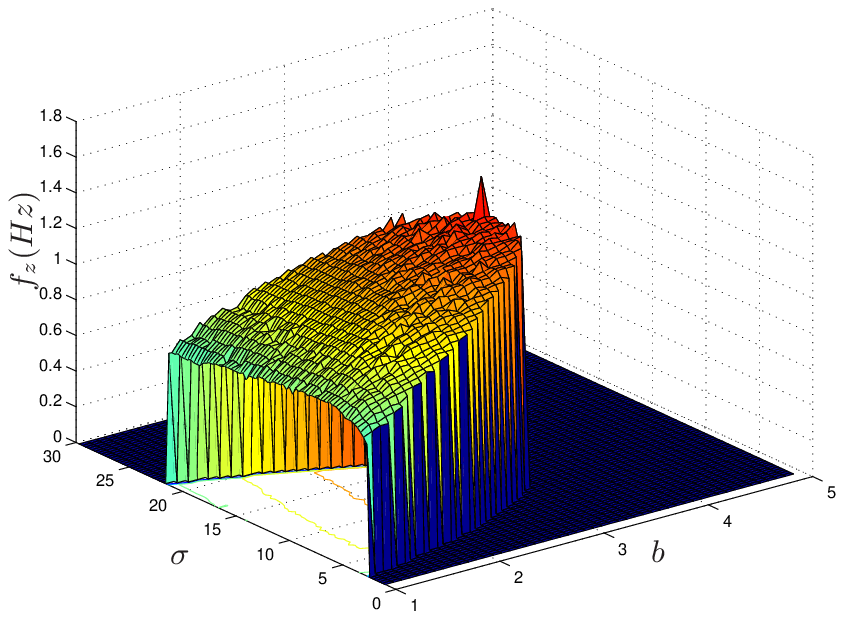}
\caption{The relation between $f_z$ and $(\sigma,b)$ when $r=28$
(some fluctuations may be induced by numerical
errors)}\label{figure:Relation_fz_r28}
\end{figure}

A plenty of experiments were performed to verify the existence of
this nearly-invariant spectral peak for other valid parameters of
the Lorenz chaotic attractor. It is found that such an inherent
frequency $f_z$ always exists and the spatial decomposition works
well. Then, what is the relation between $f_z$ and the three
parameters $r,\sigma,b$? When $r=28$, the surface of
$f_z=F(\sigma,b)$ is plotted in Fig. \ref{figure:Relation_fz_r28}.
As $r$ increases, the height of the surface will rise but the
basic shape of the surface remains. At present, an explicit
mathematical formula has not been found to describe such a
relationship.

Is it possible to extend the above results on the Lorenz attractor
to other 3-D chaotic attractors? For Chua's chaotic attractor
generated with the following equations:
\begin{eqnarray}
\dot{x} & = & p(-x+y-f(x)),\\
\dot{y} & = & x-y+z,\\
\dot{z} & = & -qy,
\end{eqnarray}
where $f(x)=m_0x+0.5(m_1-m_0)(|x+1|-|x-1|)$, our experiments show
that its inherent frequency also exists and the above spatial
decomposition works as well. Further experiments have shown the
existence of the inherent frequencies in some other 3-D
Lorenz-like chaotic attractors with double scrolls. Considering
prominent spectrum peaks also exist in many other chaotic
attractors, such as the well-known phase-coherent R\"{o}ssler
attractors
\cite{Farmer:PhaseCoherence:ANYAoS80,Anishchenoko:FreqPS:JCTE96},
it is an interesting question to ask if the spatial
decomposition-based inherent frequencies exist in \textbf{all} 3-D
chaotic attractors and how to explain this delicate phenomenon
theoretically. In other words, the question raised is: does the
only one positive Lyapunov exponent mean ``3-D chaos $=$ 1-D chaos
$+$ 2-D near periodicity"? No definite answer is given at this
time.

Finally, some possible applications of the inherent frequency and
the spatial decomposition are discussed.

\paragraph{System Identification}

The deterministic relationship between the inherent frequency
$f_z$ and the system parameters reveals a new way of realizing
system identification from the short-time waveform of any one of
the three variables $x(t)$, $y(t)$ or $z(t)$. This is useful for
adaptive synchronization of chaos
\cite{Chen-Dong:Chaos2Order1998}.

\paragraph{Chaotic Cryptanalysis}

The identification of system parameters from the inherent
frequency actually means the breaking of some chaos-based secure
communication systems, if the system parameters serve as the
secret key for encryption. In addition, another possibility of
using the inherent frequency in cryptanalysis is direct extracting
the plain-signal from the cipher-signal. For example, in chaotic
modulation systems, the plain-signal is used to change the system
parameters dynamically, which will change the short-time periods
of the transmitted cipher-signal. By distinguishing the change of
the short-time period, it is possible to directly estimate the
plain-signal without identifying the system parameters
\cite{Alvarez-Li04}.

\paragraph{Pseudo-Random Numbers Generation}

The wide-band spectrum of $ZC(i)$ means that good pseudo-random
numbers may be generated using $ZC(i)$. A possible algorithm is to
generate a symbolic 0-1 bit sequence according to which scroll the
chaotic trajectory stays in \cite[Chap. 6]{Sparrow:Lorenz1982}.
Initial experiments show that the generated pseudo-random bits can
pass some statistical tests.

\paragraph{Suppressing Chaos}

Due to the symmetry of the two sub-attractors, the Lorenz
attractor can merge into an approximately periodic single-scroll
attractor with the following transform: $x(t)\to G_x(x(t))$ and
$y(t)\to G_y(y(t))$, where $G_x(\cdot)$ and $G_y(\cdot)$ are even
symmetric functions. Such a transform can fold the two scrolls
from two different quadrants into the same quadrant and merge them
into a single scroll. As a natural result, the strongly chaotic
zero-crossing actions are suppressed and the whole attractor
becomes approximately periodic at the inherent frequency $f_z$.
Apparently, the simplest transformation functions are
$G_x(x(t))=|x(t)|$ and $G_y(y(t))=|y(t)|$, which correspond to the
merging of the positive and the negative sub-attractors obtained
via the spatial decomposition.

\begin{acknowledgments}
This research was supported by the Applied R\&D Center, City
University of Hong Kong, Hong Kong SAR, China, under Grants no.
9410011 and no. 9620004, and by the Ministerio de Ciencia y
Tecnolog\'{\i}a of Spain, research grant TIC2001-0586 and
SEG2004-02418.
\end{acknowledgments}

%\nocite{*}
\bibliographystyle{apsrev}
\bibliography{ref}

\end{document}